\documentclass[a4paper,11pt]{article}
\pdfoutput=1
\usepackage{jinstpub}
\usepackage{graphicx}
\usepackage[T1]{fontenc}
\usepackage{booktabs}
\usepackage{caption}
\usepackage{subcaption}

\title{\boldmath Jet-Parton Assignment in $t\bar{t}H$ Events using Deep Learning}

\author{M. Erdmann}
\author{B. Fischer}
\author{and M. Rieger}
\affiliation{Physics Institute 3A, RWTH Aachen University, Otto-Blumenthal-Str., 52056 Aachen, Germany}

\emailAdd{erdmann@physik.rwth-aachen.de}

\abstract{
The direct measurement of the top quark-Higgs coupling is one of the important questions in understanding the Higgs boson.
The coupling can be obtained through measurement of the top quark pair-associated Higgs boson production cross-section.
Of the multiple challenges arising in this cross-section measurement, we investigate the reconstruction of the partons originating from the hard scattering process using the measured jets in simulated $t\bar{t}H$ events.
The task corresponds to an assignment challenge of $m$ objects (jets) to $n$ other objects (partons), where $m\ge n$.
We compare several methods with emphasis on a concept based on deep learning techniques which yields the best results with more than $50\%$ of correct jet-parton assignments.
}

\keywords{Analysis and statistical methods, Computing, Data processing methods}

\arxivnumber{1706.01117} 

\begin{document}
\maketitle
\flushbottom

\section{Introduction}
\label{sec:intro}

A direct measurement of the Higgs boson coupling to top quarks is considered one of the most important consistency tests of the Higgs particle discovered in 2012 within the Standard Model of particle physics \citep{higgs_discovery,higgs_coupling}.
In this context, the mass-dependent coupling of the Higgs to matter particles and gauge bosons is expected to be largest for top quarks and close to unity.

A promising process to measure the strength of the Higgs-top coupling is to determine the cross section of top quark pair-associated Higgs boson production ($t\bar{t}H$) with the Higgs boson decaying into two bottom quarks and the top quarks decaying semi-leptonically.
The expected cross-section is sufficiently large to be detected with the LHC experiments using the data recorded thus far.
The analysis is challenged by an irreducible background from top quark pair production with radiated gluons that split into pairs of bottom quarks.
A combination of advanced analysis methods is required to separate the $t\bar{t}H$ signal from background processes.

In this work we investigate a central processing step of the analysis, namely the assignment of the jets detected in a detector to the partons of the underlying hard scattering process.
Correct assignments of the jets to the Higgs boson and to the top quarks enable calculation of suitable observables such as invariant masses and increase the separation power of signal and background.
We compare several methods to determine the jet-parton assignment with emphasis on a method based on deep learning techniques.

The application of deep learning methods in various areas of fundamental research is receiving increasing attention.
This is motivated by the impressive successes of the deep learning ansatz in handwriting recognition \citep{Hinton,Ciresan}, speech recognition \citep{Yu}, challenges with humans regarding image identification \citep{ILSVRC2012,He2015}, and in playing games \citep{Go}.
For a review, see \citep{review}.

Deep neural networks have recently been investigated for challenges in particle physics.
Various network designs have been successfully applied to extract a simulated, new exotic particle or Higgs boson signal from a background-dominated data sample \citep{exotics,higgs,kaggle}, to identify the underlying parton flavor of a jet or to measure jet substructure \citep{jetflavor,Baldi:2016fql}, and to reconstruct the neutrino flavor in neutrino-nucleus interactions \citep{NoVA}.

As a new application category of deep neural networks, we investigate the capability to select the single correct assignment from a number of possibilities of associating $m$ objects with $n$ other objects ($m\ge n$).

\begin{figure}[t!bpp]
\centering
\includegraphics[width=.57\textwidth]{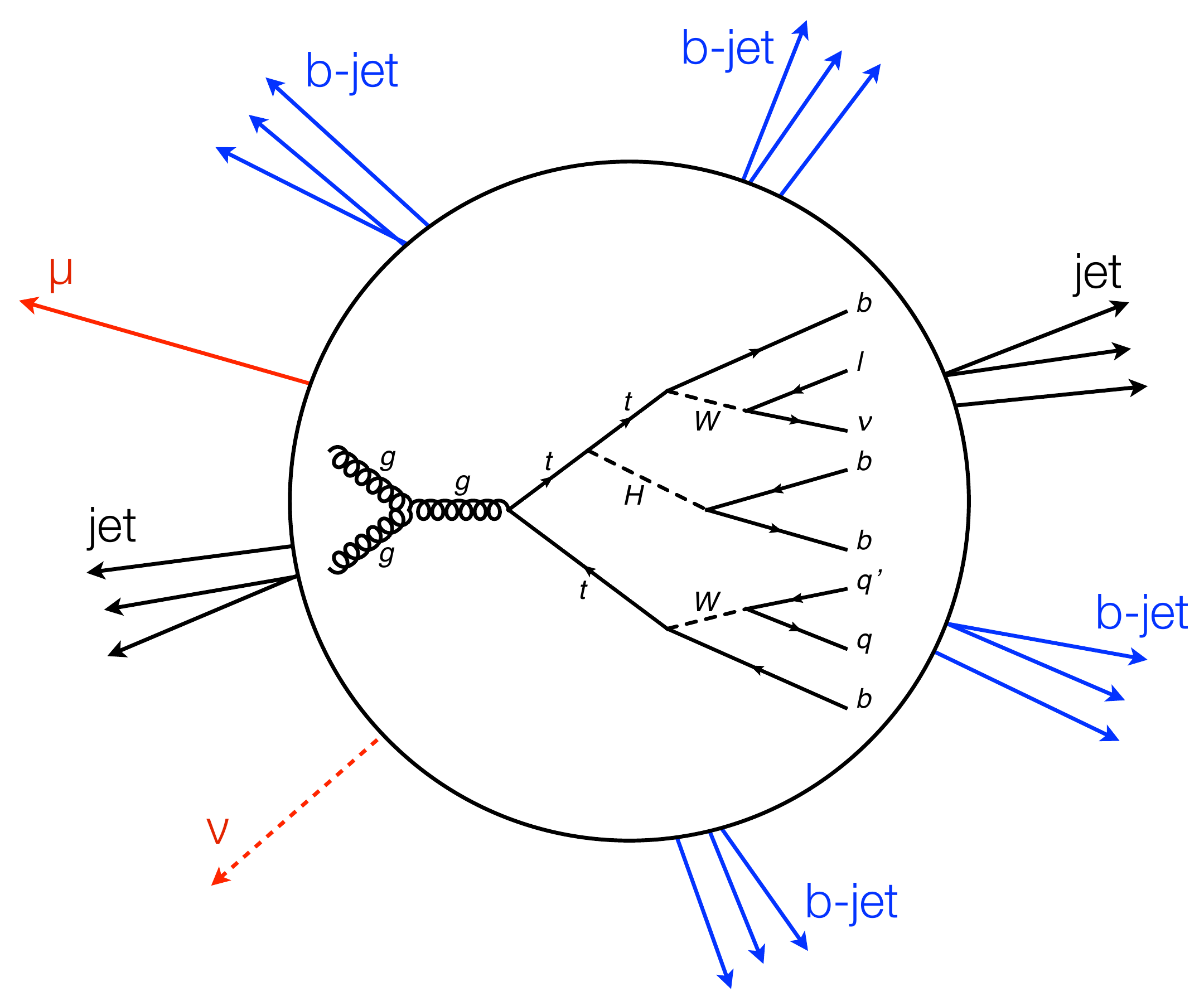}
\hfill
\caption{\label{fig:assignment} Jet-parton assignment challenge in top pair-associated Higgs
production processes ($t\bar{t}H$).}
\end{figure}

As we investigate the Higgs boson decaying to bottom quark pairs and top quark pairs in the semi-leptonic decay channel, the desired final state consists of the lepton and neutrino of one of the $W$ boson decays, two light quark jets of the other $W$ decay, and at least four bottom quark jets ($b$-jets).
A typical event situation is shown in Figure~\ref{fig:assignment}.
Note that the $b$-jets of the top and Higgs decays have similar transverse momenta such that simple kinematic cuts do not yield promising results.

The challenge thus consists of assigning $6$ out of $m\ge6$ jets to $n=6$ partons.
Higher-order radiation processes cause the number of jets to exceed $m=6$.
We reduce the number of possible jet-parton assignments (permutations) 
by ignoring the interchange of the two jets associated with the Higgs decay or the $W$ decay, respectively.
We also take advantage of bottom quark identification algorithms ($b$-tag) 
with high efficiency for correct $b$-tags and only 
a small probability of light quark jets being incorrectly $b$-tagged.

For events with $6$ jets where exactly $4$ jets are $b$-tagged, we assign the two untagged jets
to the $W$ boson and obtain $4!/2=12$ permutations for the $b$-tagged jets.
Only one of the $12$ permutations corresponds to the correct jet-parton assignment.
Owing to higher-order radiation effects with correspondingly more jets observed in the detector, the majority of events allows for a larger number of possible permutations which typically reaches several hundred.

We will evaluate the efficiency of the deep neural network to find the correct permutations and compare them with the results of a boosted decision tree and a $\chi^2$ measure derived from the hypothetical masses of a permutation.

This work is structured as follows:
Initially, the deep network design is explained.
The simulated dataset is then introduced and input observables are presented in the following sections.
We evaluate the efficiencies of finding the correct assignment with different methods.
We also demonstrate the effect on reconstructed observables which are typically used in $t\bar{t}H$ analyses.
Finally, we present our conclusions.

\section{Neural network design}
\label{sec:network}

\begin{figure}[t!bp]
\captionsetup[subfigure]{aboveskip=-1pt,belowskip=-1pt}
\begin{centering}
\begin{subfigure}[b]{0.25\textwidth}
\includegraphics[width=\textwidth]{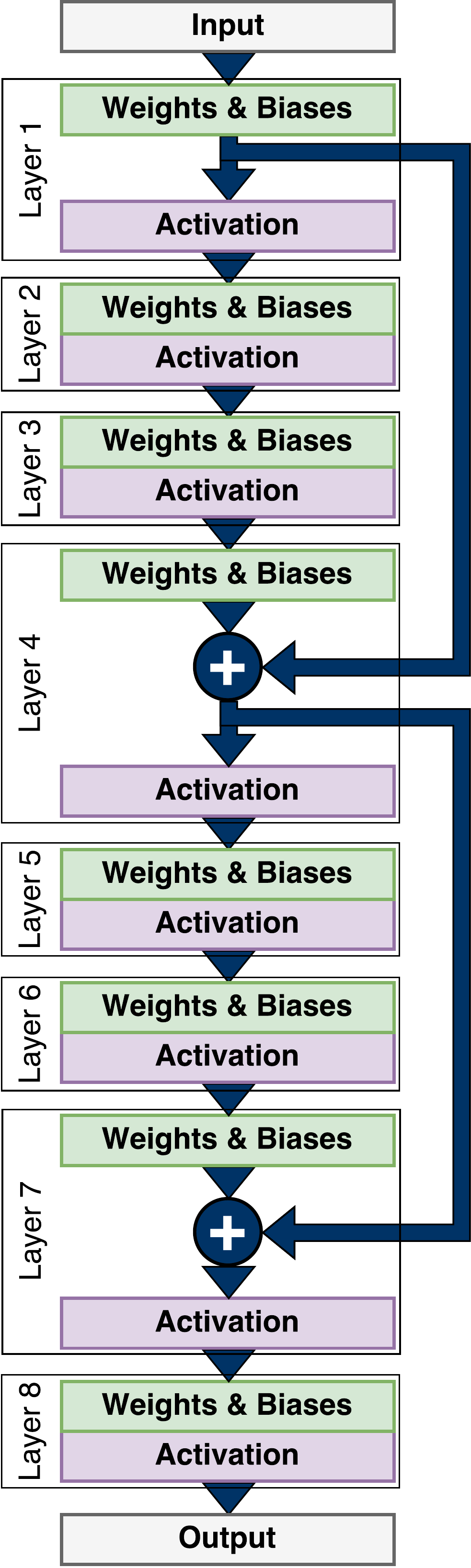}
\subcaption{}
\label{fig:architecture_nn}
\end{subfigure}
\hspace{25mm}
\begin{subfigure}[b]{0.35\textwidth}
\includegraphics[width=\textwidth]{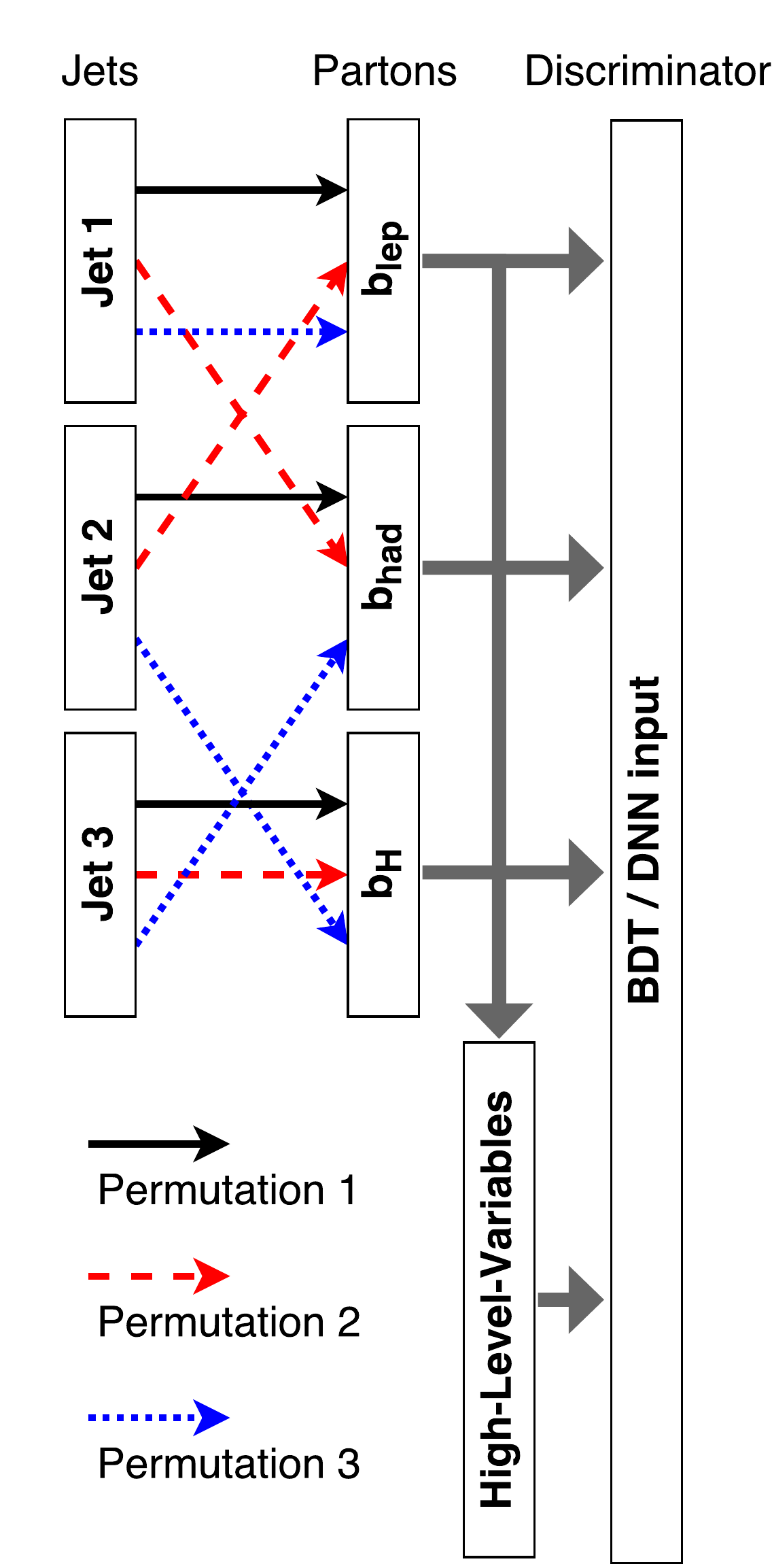}
\subcaption{}
\label{fig:architecture_perm}
\end{subfigure}
\caption{
a) Fully connected architecture of the network with additional connections following the concept of residual networks.
b) Permutations of the input variables to the neural network (DNN) or the boosted decision trees (BDT), respectively.
}
\label{fig:architecture}
\end{centering}
\end{figure}

For the network architecture we use various technical concepts developed within deep learning research.
As we work with a fixed number of features, i.e. input observables, our basic concept is a fully connected network consisting of $8$ hidden layers with $500$ nodes in each layer (Figure~\ref{fig:architecture_nn}, Table~\ref{tab:design}).

In addition, we apply the so-called residual network concept \citep{He2015} where the input to the layer $i$ not only consists of the output of the previous layer $i-1$, but also from another previous layer $i-j$, where we choose $j=3$:
\begin{align}
{\rm input}_i &= {\rm output}_{i-1}+{\rm output}_{i-3}
\label{eq:residual}
\end{align}
This residual concept is found to accelerate the training of the deep neural network.
For modeling non-linearities, we apply an exponential linear (ELU) activation function on the output of each node.

The network is trained to find the best match between the partons of the fundamental hard scattering process and the final state particles as measured in a collider detector.
In Figure~\ref{fig:architecture_perm} we exemplarily show input variables and illustrate the training principle.
The partons $b_{\rm lep}, b_{\rm had},...$ and the high level variables calculated from them define the input layer of the network.
A correct jet-parton assignment is represented by ${\rm jet}_1$ being the $b$-jet truly originating from the bottom quark $b_{\rm lep}$ of the leptonic top quark decay,
${\rm jet}_2$ truly originating from the bottom quark $b_{\rm had}$ of the hadronic top quark decay, etc.
For the incorrect jet-parton assignments, the order of the jets on input is changed accordingly.

The cost function to be minimized in the training process is described by the binomial cross-entropy:
\begin{align}
J(W,b \vert \hat{D},D_{\rm DNN}(W,b,x)) = - \left[ \hat{D}\cdot \log{(D_{\rm DNN})} + \omega\cdot \left(1 - \hat{D}\right)\cdot \log{(1 - D_{\rm DNN})}\right] + \lambda \sum_j^N W_j^2
\label{eq:loss_function}
\end{align}
Here, $D_{\rm DNN}$ is the output of the neural network given the parameters $W$ and $b$ for input observables $x$, which contain all information on the final state particles.
$\hat{D}$ is the training target whose value is $1$ for correct jet-parton assignments and $0$ otherwise.
As the number of incorrect permutations by far exceeds the number of correct ones, incorrect permutations are scaled by $\omega$ to give the same weight in the training.
Below we will apply a preselection of at most $50$ permutations such that the weights for incorrect permutations
vary within $\omega\in [1/12,1/50]$. 

During training, the weights $W$ of the network and the biases $b$ are adjusted such that the network returns the result of a logistic function as the discriminator $D_{\rm DNN}\,\epsilon\,[0,1]$ which scores a permutation given the input $x$.
For all possible assignments within one event, the trained discriminator of a perfectly trained network should give the maximum discriminator $D_{\rm DNN}=D_{max}$ for the correct jet-parton assignment, and a smaller value $D_{\rm DNN}<D_{max}$ for all other permutations.
A set of $10^{4}$ permutations $(x_i,\hat{D}_{i})$ forms a training batch, randomly selected from all simulated collision events.

In order to ensure generalization and reduce prediction errors when applied to a new, independent dataset, we introduce regularization measures.
We use $L^2$ regularization (last term in eq.(\ref{eq:loss_function})) which penalizes large weights during training and prevents the network from learning features unique to the training set.

We initialize the weights $W$ according to a Gaussian distribution with squared width $\sigma^2=2/(l_{in}+l_{out})$ where $l_{in}$ and $l_{out}$ are the number of ingoing and outgoing connections of a layer \citep{Glorot}.
We also pre-process the input observables using feature scaling where we normalize $x$ to mean $0$ and standard deviation $1$.
For the optimization we use the concept of the stochastic gradient descent with an adaptive learning rate (ADAM) \cite{ADAM}.

We use the TensorFlow software package to set up the network \cite{tensorflow}.
The network training is performed on a cluster with GPU cards of GeForce GTX 1080 type with up to 64 GB of system memory.
A typical training time amounts to $6$~h.

\begin{table}[t!bp]
\begin{center}
\begin{tabular}{ll}
\toprule
Parameter & Value \\
\midrule
Hidden layers & 8 \\
Units per layer & 500 \\
Activation & ELU \\
$L^2$ factor ($\lambda$) & $4 \cdot 10^{-6}$ \\
Batch size & 10,000 \\
Optimizer & ADAM \\
Learning rate & 0.001 \\
Training epochs & 200 \\
\bottomrule
\end{tabular}
\caption{Design parameters of the neural network.}
\label{tab:design}
\end{center}
\end{table}

\section{Simulated dataset}

To simulate $t\bar{t}H$ events we use the Pythia $8.2.19$ program package \citep{Pythia}.
The matrix elements for $t\bar{t}H$ processes include angular correlations of the decay products from heavy resonances.
The beam conditions correspond to LHC proton-proton collisions at $\sqrt{s}=13$~TeV.
We use only the dominant gluon-gluon process and the Higgs boson decay into a bottom quark pair.
Hadronization is performed with the Lund string fragmentation concept.

In order to analyze a typical final state as observed in a LHC detector we use the DELPHES package to simulate the CMS detector \citep{deFavereau:2013fsa}.
The DELPHES project provides a modular framework that simulates a multipurpose detector in a parameterized way.
It includes all major effects such as pile-up, charged particle deflections in magnetic fields, electromagnetic and hadronic calorimetry, and muon detection systems.
The simulated output consists of muons, electrons, photons, jets and missing transverse momentum from a particle flow algorithm as well as identifiers for bottom jets and
tau leptons.

For the jet finding the anti-$k_t$ algorithm is used with the size parameter $R=0.5$.
For the $b$-tag algorithm both the efficiency $\varepsilon_b$ of correct $b$-tags
and the probability $\rho_b$ of jets from light quarks being incorrectly $b$-tagged 
have a dependency on the jet transverse momentum $p^{jet}_t$ with the values
$\varepsilon_b=0.71$ and $\rho_b=0.014$ at $p^{jet}_t=100$~GeV.

In our analysis of the simulated data, we select events with lepton transverse momenta above $p^l_{t}=20$~GeV (electrons or muons) and their pseudorapidities within $\vert \eta^l \vert < 2.1$.
The jet transverse momenta are above $p^{jet}_{t}=25$~GeV, with their pseudorapidities being within $\vert \eta^{jet} \vert < 2.5$.
We require the number of jets to be within $6\le n^{jet}\le 10$; of these at least $n^{jet}=2$ must be $b$-tagged.

The correct assignment is obtained by matching partons with jets using generator information. A match must fulfill $\Delta R_{jet,parton}<0.3$ with
\begin{align}
\Delta R = \sqrt{\Delta \phi^2 + \Delta \eta^2}\;\;,
\end{align}
where $\phi$ is the azimuthal angle in the detector and $\eta$ denotes the pseudorapidity.
Ambiguities are resolved by minimizing the sum $\sum{\Delta R}$ of all viable matches.
In order to define valid training targets, we consider only events for which all partons could be matched.

From the total of $10^{8}$ generated events, 
we could identify all partons of the hard scattering process in $44\%$ of the events.
After applying the selection criteria for reconstructed jets and leptons $5.6\%$ of the generated events remained.
The additional requirement of the above-mentioned jet-parton match was then fulfilled by $0.7\%$ of the events.
Therefore, we perform the investigations with $738,270$ events, using $40\%$ for training purposes and $10\%$ for immediate validation.
To determine the correct assignment efficiency we use $50\%$ of the events.

\section{Detector observables}
\label{sec:observables}

We use two categories of variables as the input observables for solving the jet-parton assignment in $t\bar{t}H$ events:
\begin{enumerate}
 \item Basic particle-related variables:
  \begin{enumerate}
   \item For all jets and leptons: four-vector values such as the momenta in both spherical and Cartesian coordinates.
   \item For the charged lepton: isolation variables and sum of transverse momenta $\Sigma p_{T}$ of charged and neutral particles within $\Delta R \le 0.5$ relative to the lepton momentum.
   \item For jets: b- and $\tau$-tag values, jet area, the number of constituents, and electromagnetic and hadronic calorimeter contributions.
  \end{enumerate}
 \item Observables derived from combinations of several objects:
  \begin{enumerate}
   \item The masses of the Higgs boson, the top quarks, the hadronically decaying $W$ boson, and the individual $\chi^2$ values compared to the expected masses.
   Also the combined mass of the $t\bar{t}$ system.
   \item The transverse momenta of the $t\bar{t}$ system and the $t\bar{t}H$ system, and the ratio of the jet scalar transverse momentum sum to that of all final state particles including the neutrino.
   \item The distance $\Delta R$ between the two jets originating from the Higgs boson decay, and their differences in pseudorapidity and in azimuthal angles.
   The distance $\Delta R$ between the two jets originating from the $W$ boson decay, and their differences in pseudorapidity and in azimuthal angles.
   The difference between the pseudorapidities of the lepton and the $b$-jet of the leptonically decaying top quark.
   \item The spatial angle $\theta^*$ between the charged lepton in the $W$ boson rest frame and 
   the $W$ boson direction when boosted into the rest frame of its corresponding top quark.
   The spatial angle $\theta^*$ between the softer jet of the $W$ boson decay in the $W$ boson rest frame and the $W$ boson direction
   in the rest frame of its corresponding top quark.
  \end{enumerate}
\end{enumerate}
The neutrino was reconstructed using the charged lepton, and the missing transverse energy according to ~\citep{nureco}.

\begin{figure}[t!bp]
\captionsetup[subfigure]{aboveskip=-1pt,belowskip=-1pt}
\begin{centering}
\begin{subfigure}[b]{0.495\textwidth}
\includegraphics[width=\textwidth]{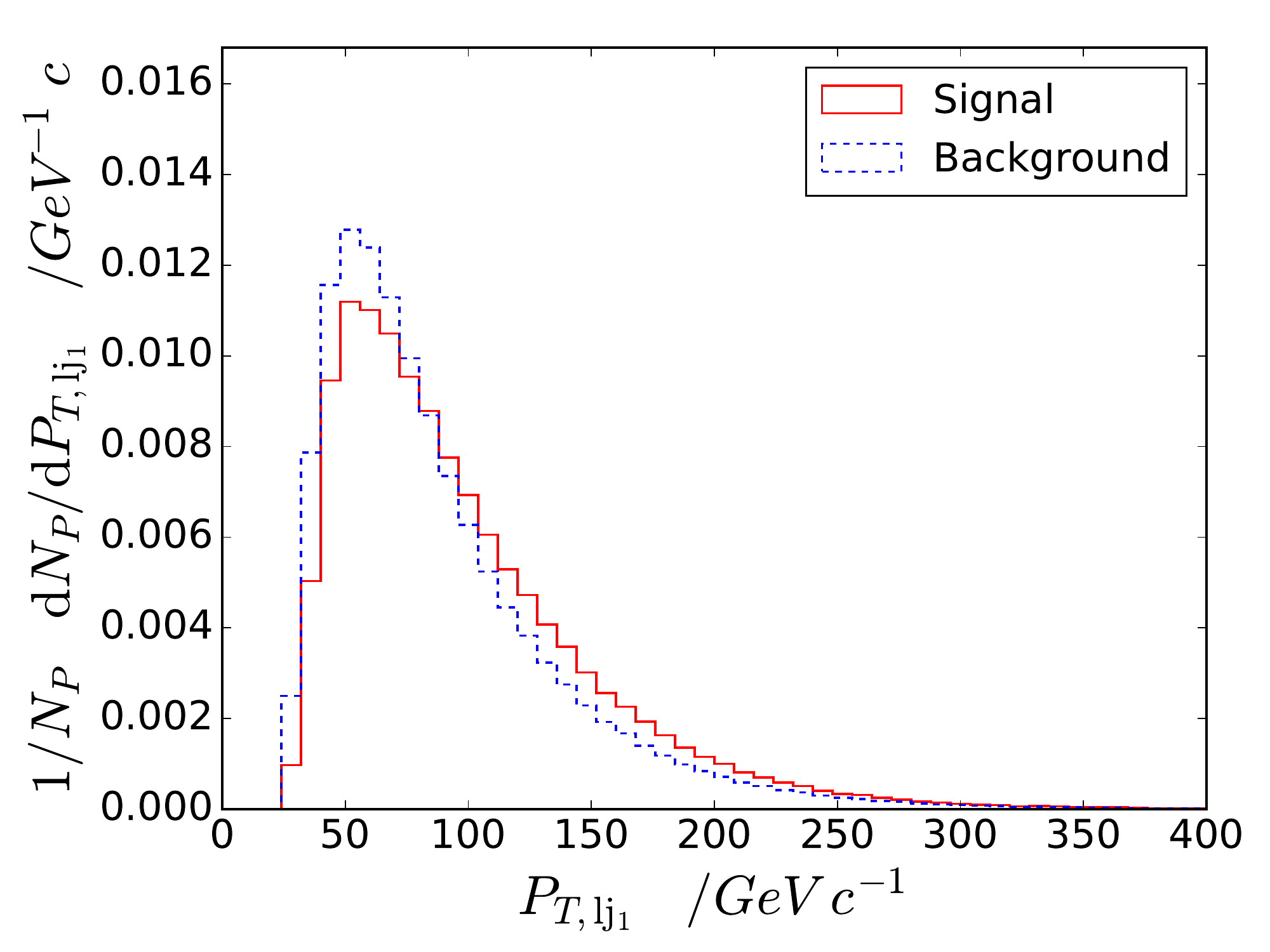}
\subcaption{}
\label{fig:observables_a}
\end{subfigure}
\hfill
\begin{subfigure}[b]{0.495\textwidth}
\includegraphics[width=\textwidth]{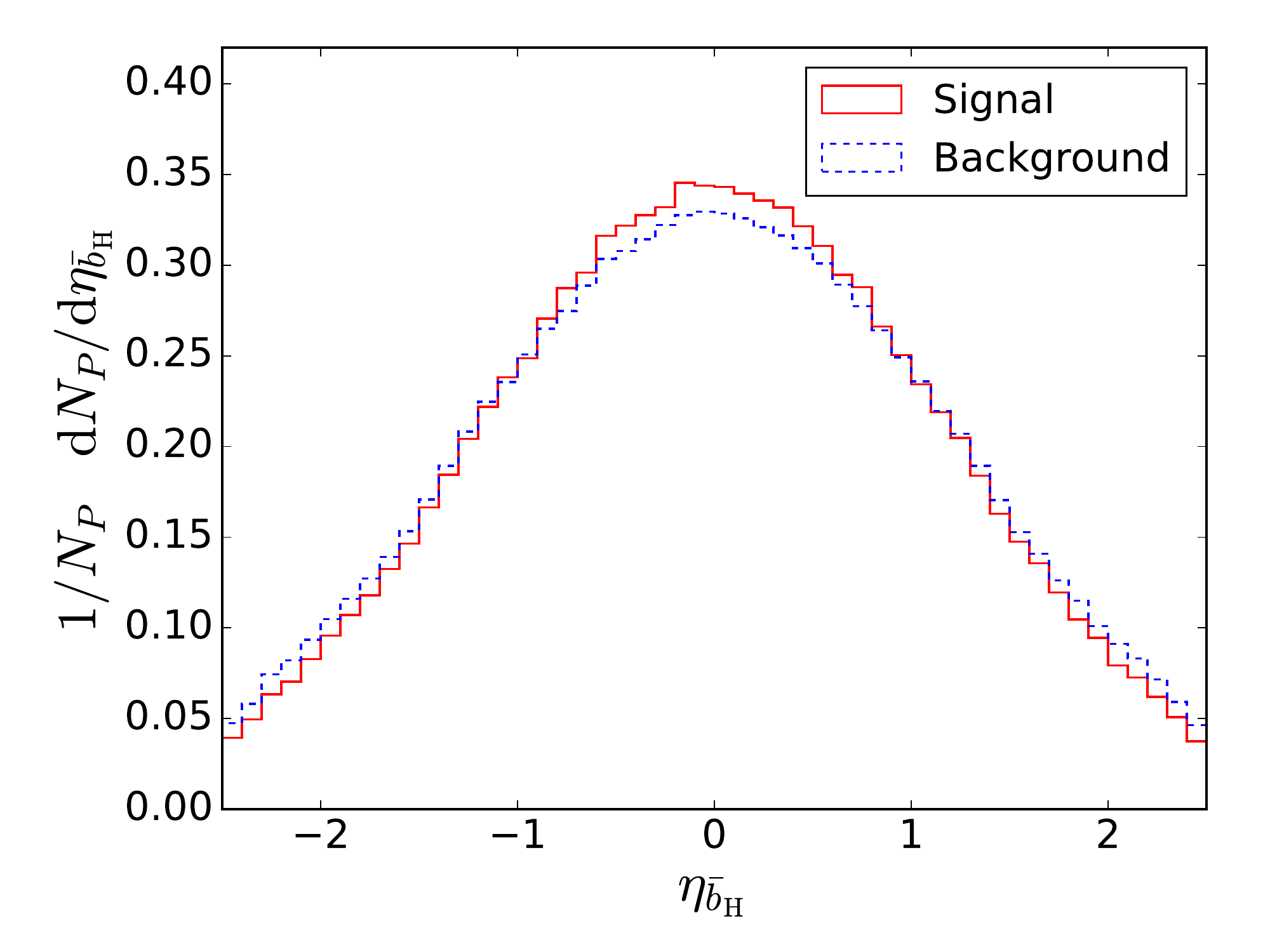}
\subcaption{}
\label{fig:observables_b}
\end{subfigure}
\begin{subfigure}[b]{0.495\textwidth}
\includegraphics[width=\textwidth]{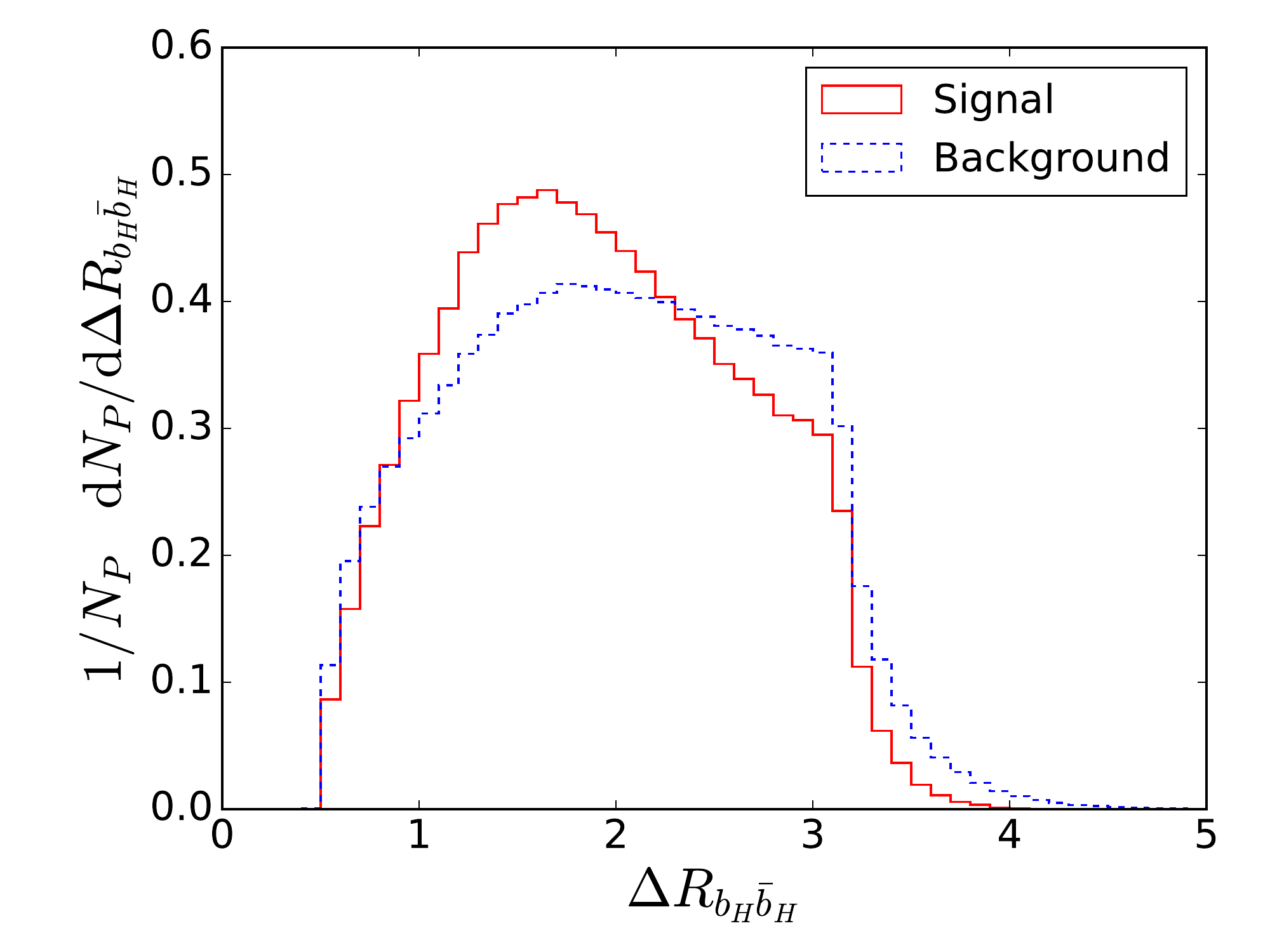}
\subcaption{}
\label{fig:observables_c}
\end{subfigure}
\hfill
\begin{subfigure}[b]{0.495\textwidth}
\includegraphics[width=\textwidth]{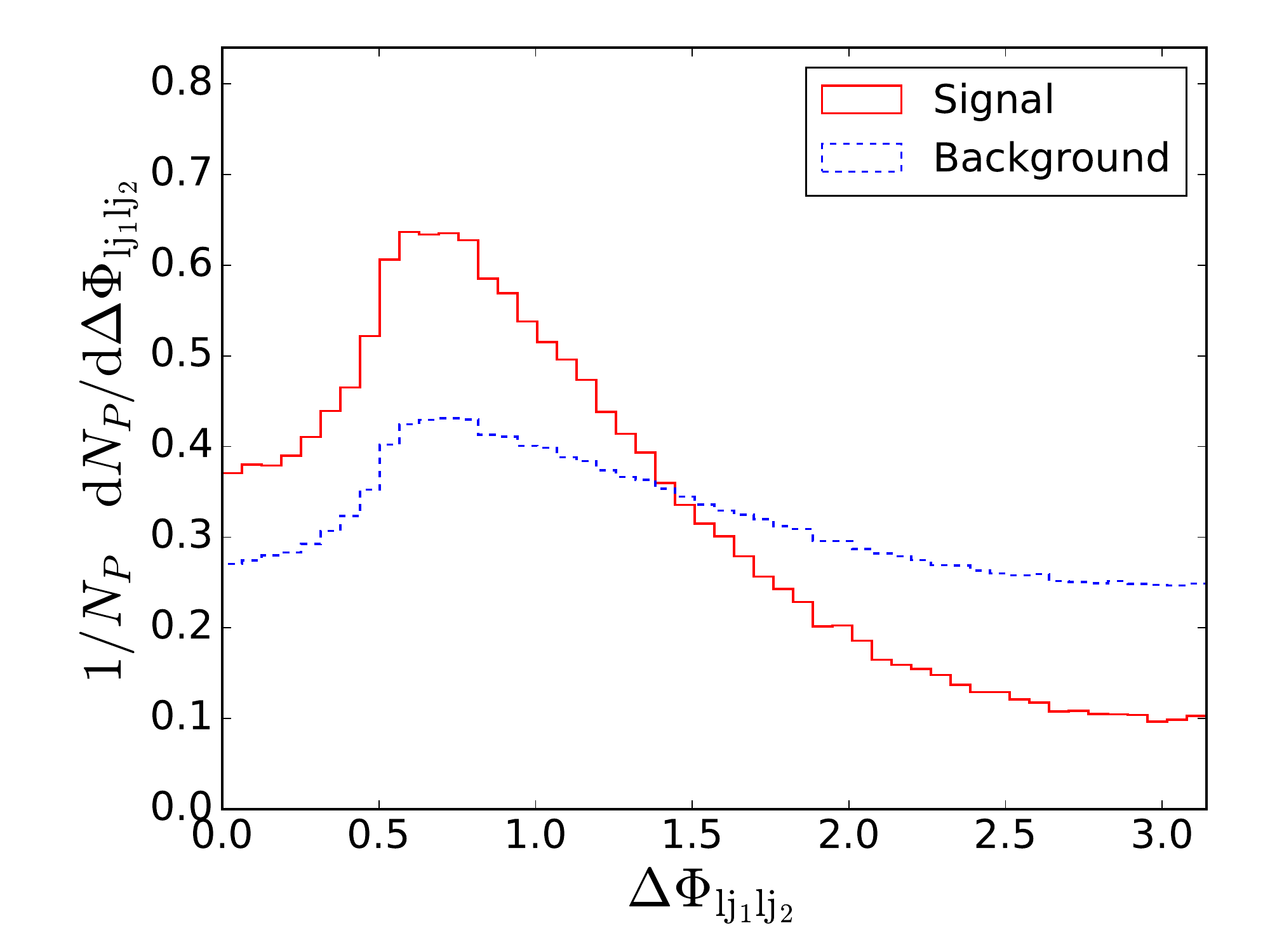}
\subcaption{}
\label{fig:observables_d}
\end{subfigure}
\begin{subfigure}[b]{0.495\textwidth}
\includegraphics[width=\textwidth]{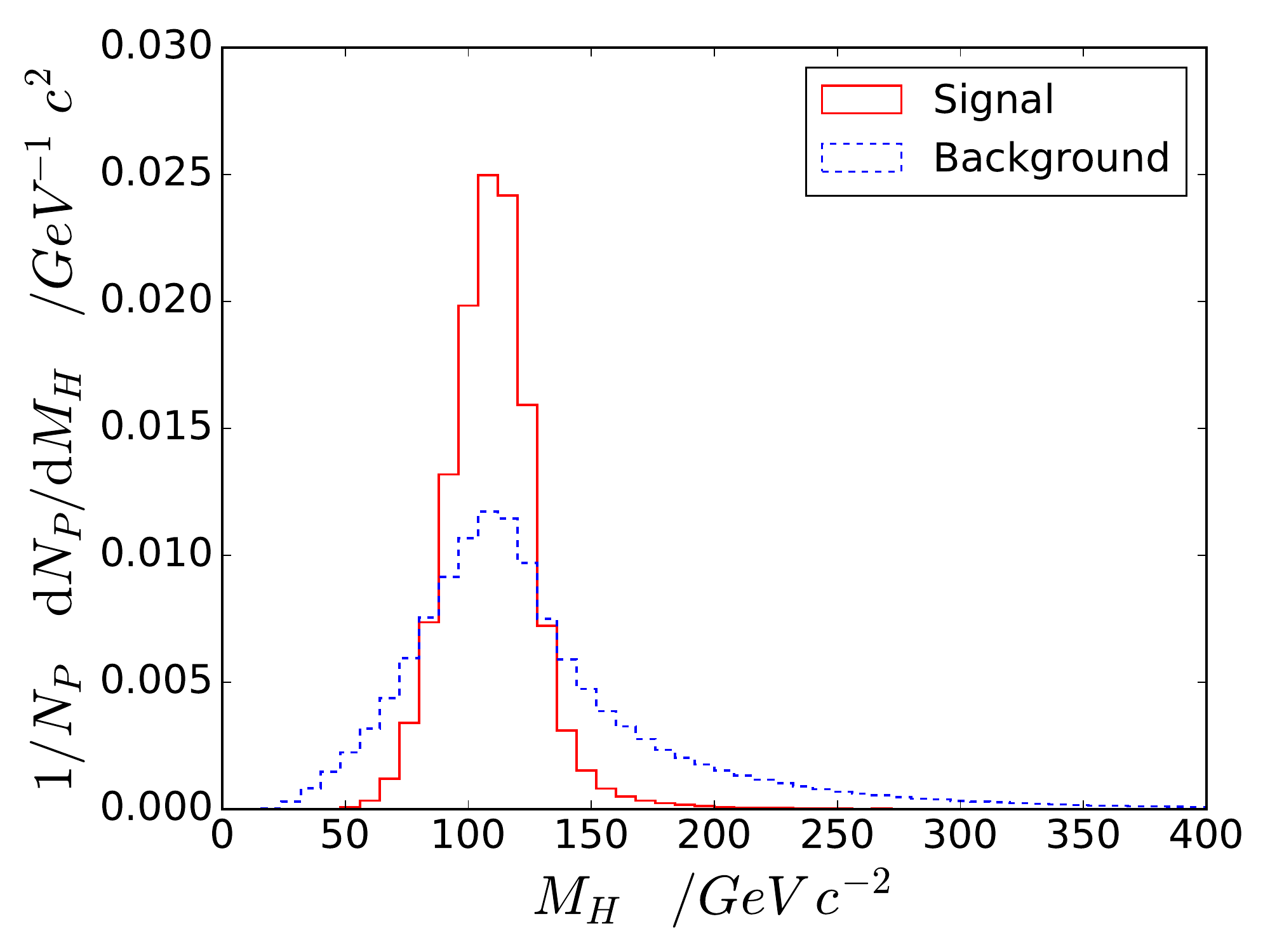}
\subcaption{}
\label{fig:observables_e}
\end{subfigure}
\hfill
\begin{subfigure}[b]{0.495\textwidth}
\includegraphics[width=\textwidth]{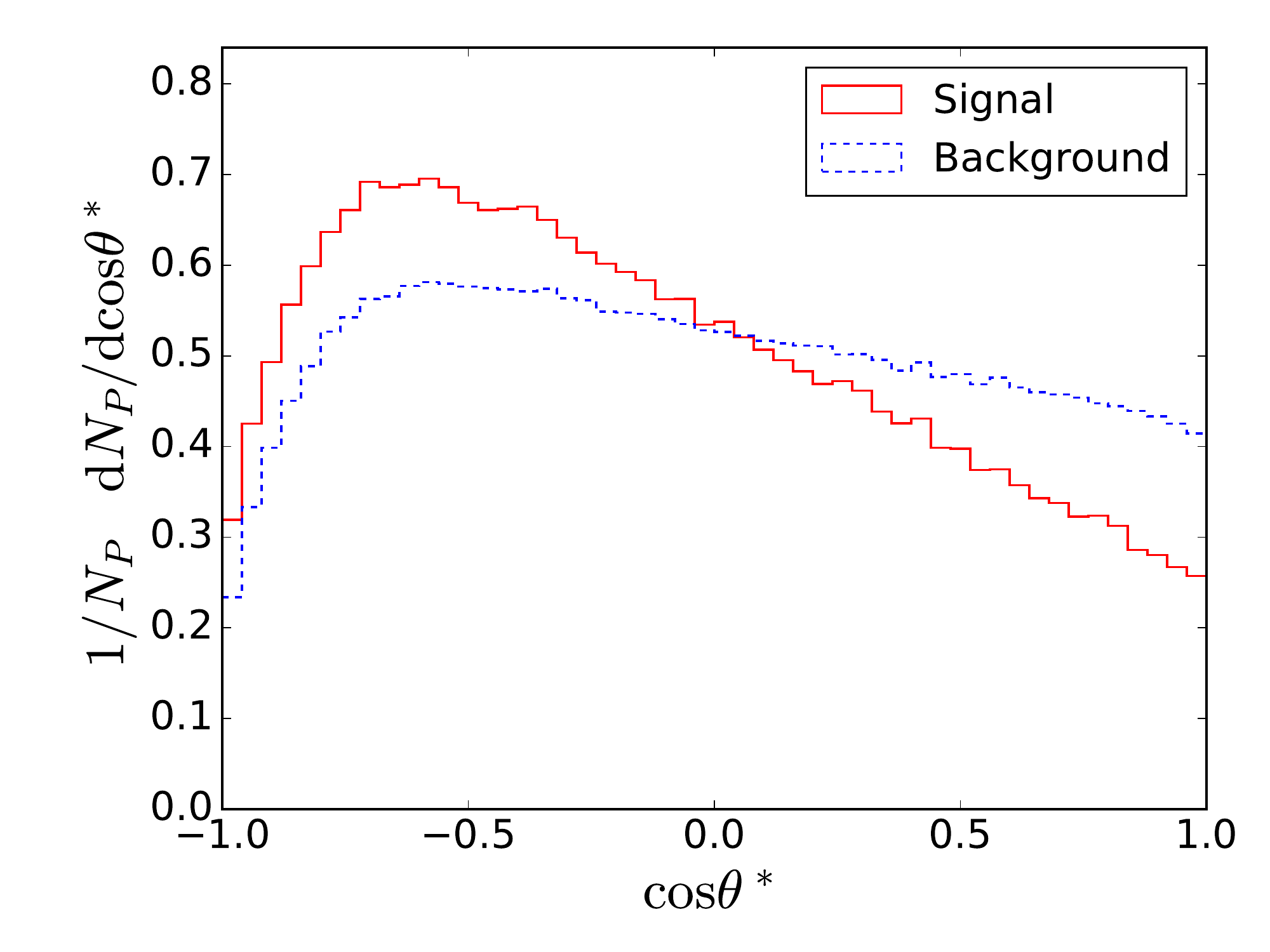}
\subcaption{}
\label{fig:observables_f}
\end{subfigure}
\caption{Observables used for the training with the correct jet-parton assignment (full red curve) and permutations (dashed blue curve): a) transverse momentum of the leading jet of the $W$ boson decay, b) pseudorapidity of the jet induced by the bottom anti-quark from the Higgs decay, c) directional distance $\Delta R$ of the jets of the Higgs boson decay, d) azimuthal angular distance of the two jets of the $W$ boson decay, e) reconstructed Higgs mass, and f) the spatial angle $\cos{\theta^*}$ of the leptonic branch.}
\label{fig:observables}
\end{centering}
\end{figure}

As an example of how different the distributions are when selecting the correct jet-parton assignment or permutations, respectively, in Figure~\ref{fig:observables_a} we show the transverse momentum distribution of the $p_t$ leading light quark jet resulting from the $W$ decay.
The full red curve represents the distribution from correct assignments of the jet to a quark from the $W$ decay.
The dashed blue curve illustrates the distribution resulting from incorrectly assigning a jet to the $W$ decay.
The combinatorics of wrong assignments by far exceeds the single true assignment solution.

In Figure~\ref{fig:observables} we show similar distributions for b) the pseudorapidity of the jet induced by the bottom anti-quark from the Higgs boson decay, c) the distance $\Delta R$ between the jets originating from the Higgs boson, d) the azimuthal angular distance of the two jets from the $W$ boson decay, e) the reconstructed mass of the Higgs boson, and f) the cosine of the spatial angle $\theta^*$ of the leptonic branch as defined in item $2(d)$ above.

In total, we use $140$ basic kinematic observables of the above first category, and $22$ combined observables of the second category.

\section{Jet-parton assignments}

The efficiencies of finding the correct jet-parton assignment using three different methods are described in the following.
As a benchmark, drawing the correct jet-parton assignments by chance is largest for events with $6$ jets where the four $b$-jets were correctly tagged: $1/12\approx 8\%$.
As the experimental $b$-tagging efficiency is $\varepsilon<1$ and the probability of successfully tagging $k=4$ $b$-jets is small, $\varepsilon^k\ll 1$, this is a rare case within our simulations.
The majority of $t\bar{t}H$ events has more than $6$ jets which results in a negligible rate of correct assignments by chance.

\subsection{$\chi^2$ method}

Initially, we use the reconstructed masses of the Higgs boson, both top quarks, and the $W$ boson of the hadronically decaying top quark in the context of a $\chi^2$ measure:
\begin{align}
\chi^2 = \left(\frac{m_H-\tilde{m}_H}{\tilde{\sigma}_H}\right)^2
       + \left(\frac{m_{t,lep}-\tilde{m}_{t,lep}}{\tilde{\sigma}_{t,lep}}\right)^2
       + \left(\frac{m_{t,had}-\tilde{m}_{t,had}}{\tilde{\sigma}_{t,had}}\right)^2
       + \left(\frac{m_{W,had}-\tilde{m}_{W,had}}{\tilde{\sigma}_{W,had}}\right)^2
\label{eq:chi2}
\end{align}
Here, the index $H$ refers to the Higgs boson, $({t,lep})$ to the top quark of the leptonic branch, and $({t,had})$ and $({W,had})$ to the top quark and $W$ boson of the hadronic branch.
The tilde symbol indicates the reconstructed average masses $\tilde{m}$ using the correct jet-parton assignments, and the corresponding widths $\tilde{\sigma}$ of the mass distributions.

\begin{figure}[t!bp]
\captionsetup[subfigure]{aboveskip=-1pt,belowskip=-1pt}
\begin{centering}
\begin{subfigure}[b]{0.495\textwidth}
\includegraphics[width=\textwidth]{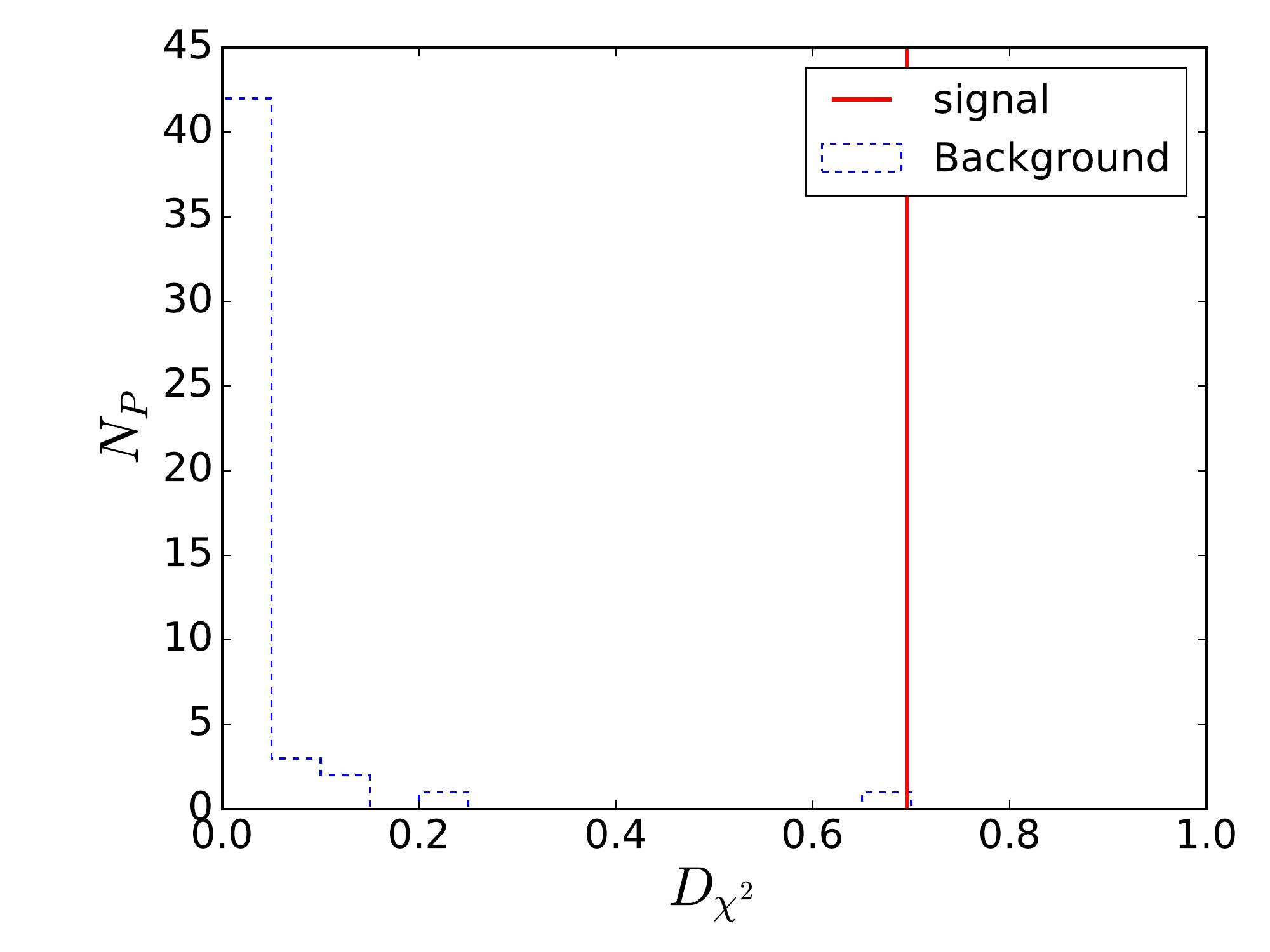}
\subcaption{}
\label{fig:score_a}
\end{subfigure}
\hfill
\begin{subfigure}[b]{0.495\textwidth}
\includegraphics[width=\textwidth]{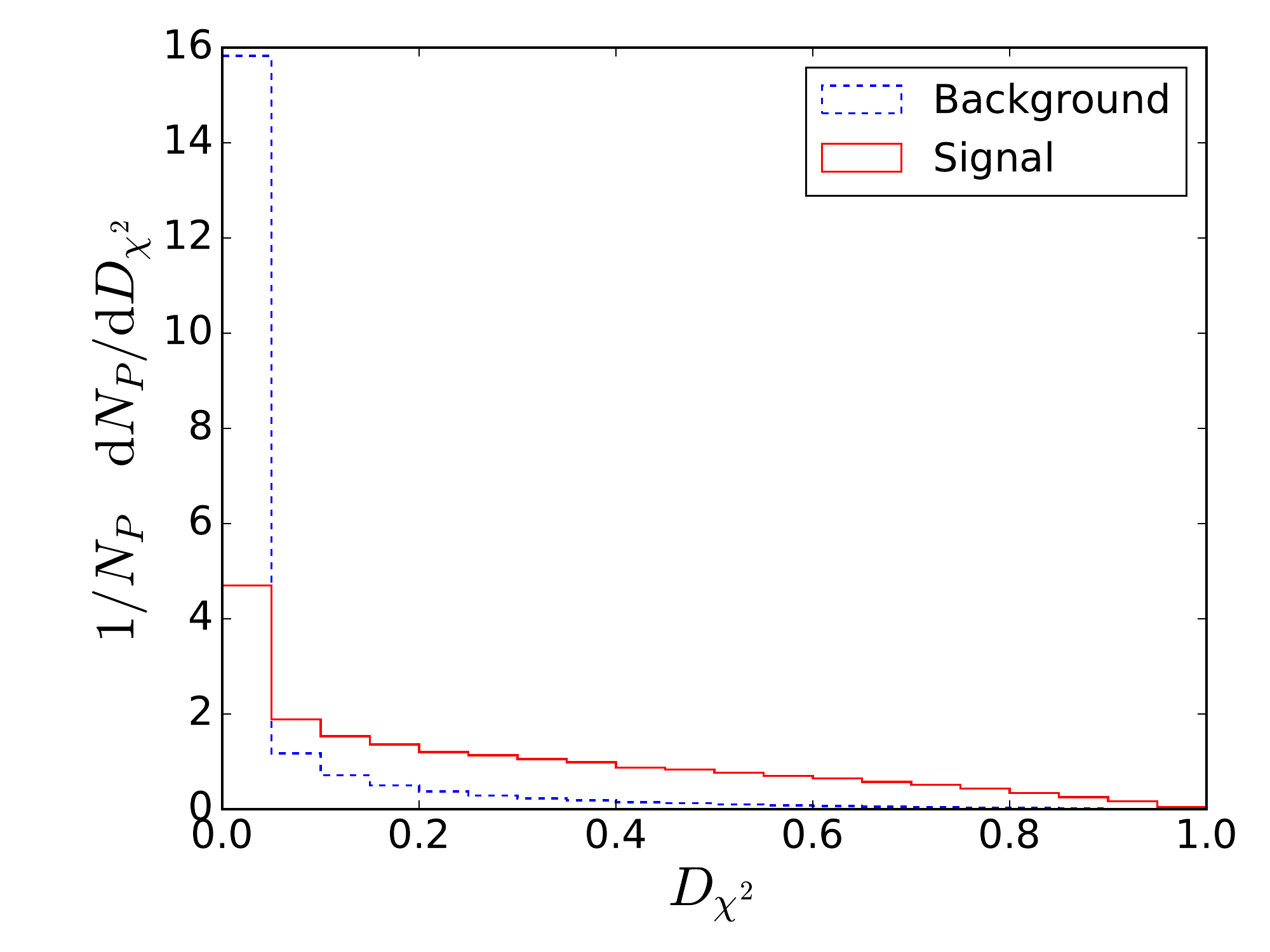}
\subcaption{}
\label{fig:score_b}
\end{subfigure}
\begin{subfigure}[b]{0.495\textwidth}
\includegraphics[width=\textwidth]{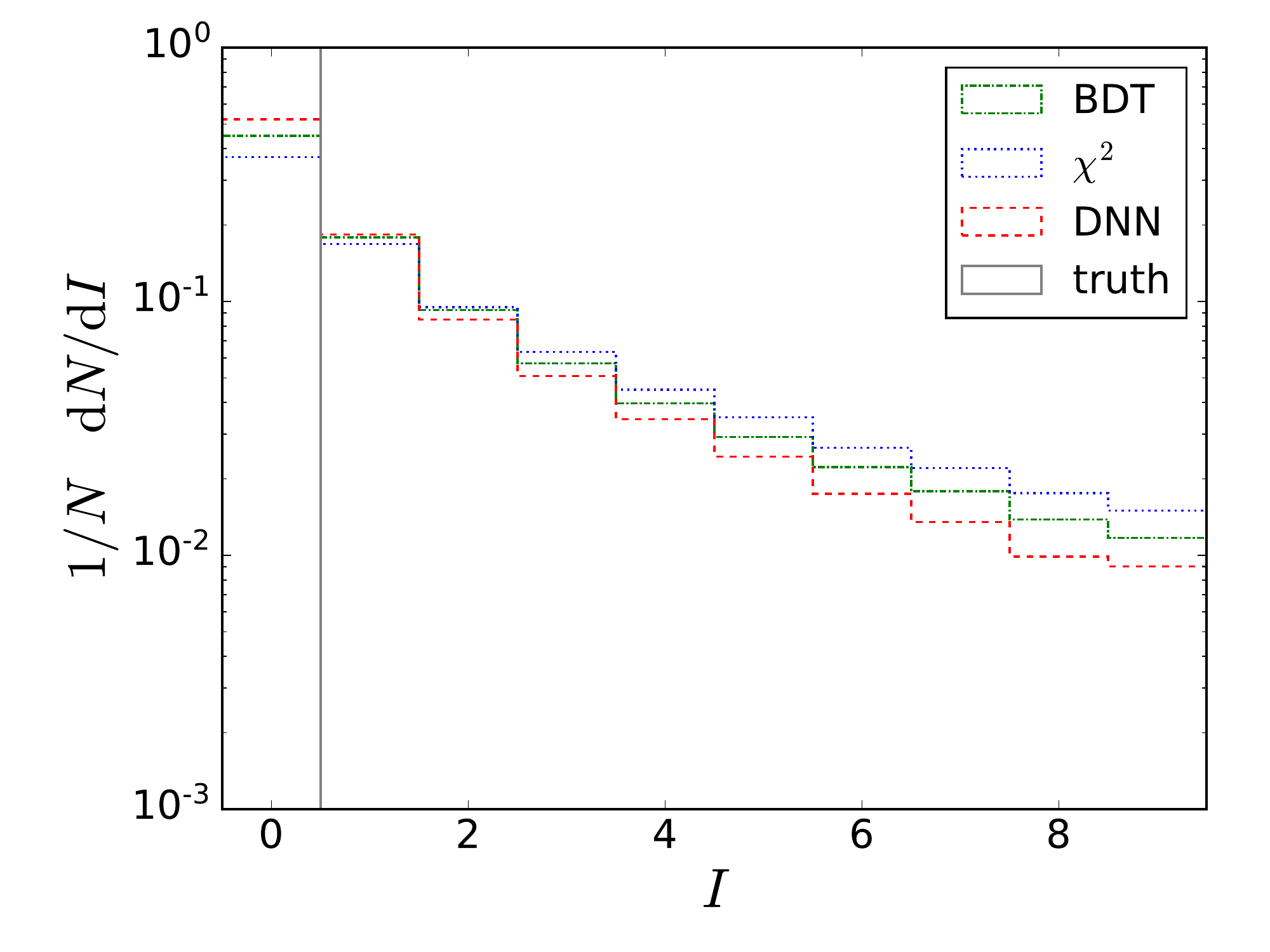}
\subcaption{}
\label{fig:score_c}
\end{subfigure}
\hfill
\begin{subfigure}[b]{0.495\textwidth}
\includegraphics[width=\textwidth]{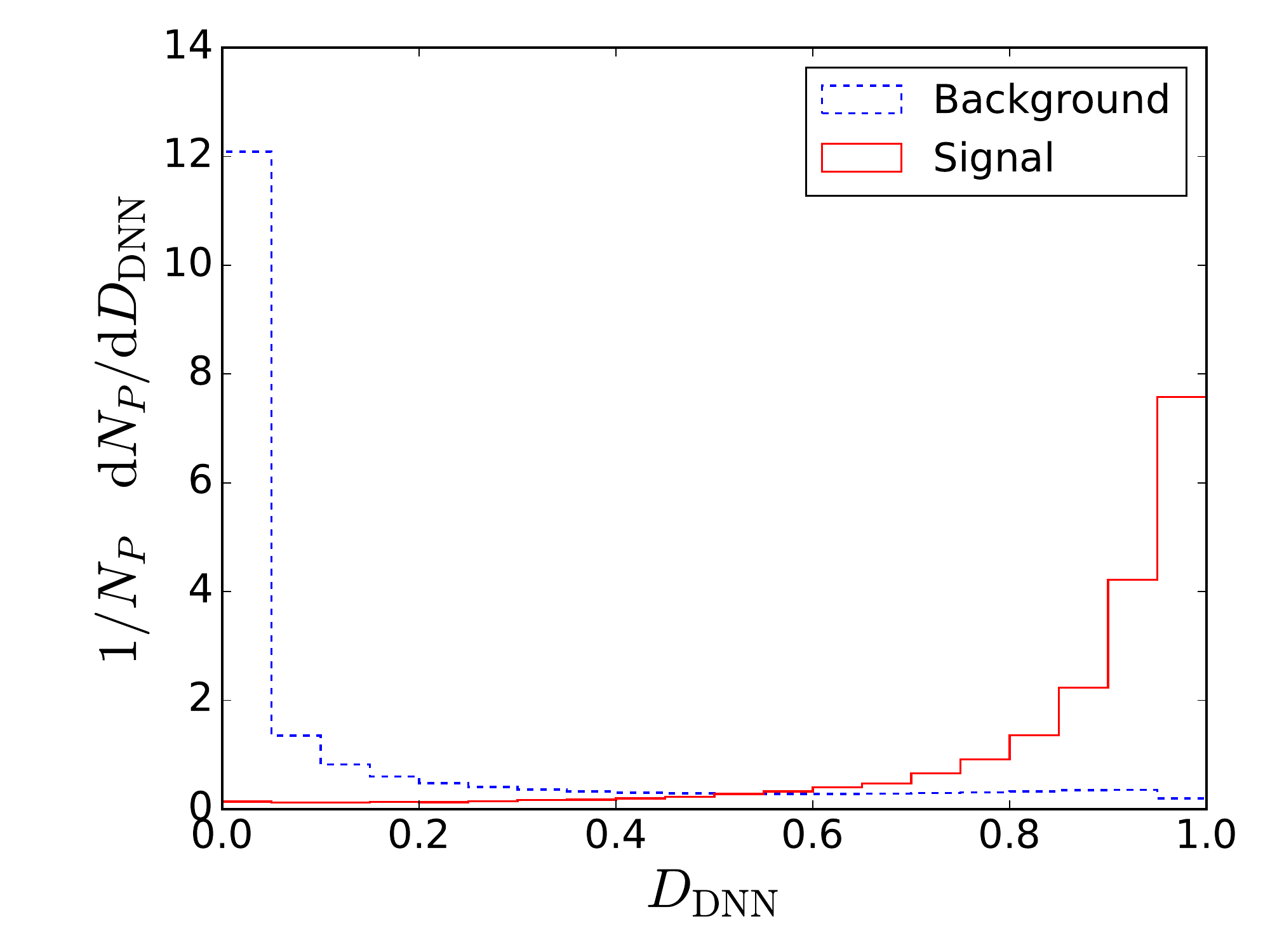}
\subcaption{}
\label{fig:score_d}
\end{subfigure}
\caption{a) Distribution of the discriminator $D_{\chi^{2}}=\exp{(-\chi^2)}$ of the $\chi^{2}$ method for one example event where the largest value is chosen as the reconstructed jet-parton assignment.
The discriminator value of the correct assignment is shown by the vertical line.
b) Distribution of the discriminator $D_{\chi^{2}}$ for all events with the correct assignment (full red curve) compared to other permutations (dashed blue curve).
c) Number of permutations with discriminator values above that of the correct jet-parton assignment as a measure of the quality of the reconstructed assignment (\ref{eq:integral}).
The full curve illustrates the correct assignments and the dashed curves the neural network (red), the boosted decision tree (green), and the $\chi^{2}$ method (blue).
d) Distribution of the neural network discriminator $D_{\rm DNN}$ for all events with the correct assignment (full red curve) compared to other permutations (dashed blue curve).}
\label{fig:score}
\end{centering}
\end{figure}

In Figure~\ref{fig:score_a} we show the distribution of our discriminator $D_{\chi^{2}}\equiv \exp{(-\chi^2)}$ for permutations of an exemplary event.
We use this definition of $D_{\chi^{2}}$ for consistency with the subsequent methods, which return better results at higher scores.
Correspondingly, the permutation with the highest $D_{\chi^{2}}$ value (smallest $\chi^{2}$) is considered the selected jet-parton assignment and eventually denotes the full event reconstruction.
The correct jet-parton assignment is typically contained within the $50$ permutations with the largest $D_{\chi^{2}}$ discriminator values.
Therefore we investigate only these $50$ permutations.

In the same figure, we show the discriminator $D_{\chi^{2}}^{truth}$ resulting from the correct jet-parton assignment as a vertical line which coincides in this event with the reconstructed assignment.
Also shown in Figure~\ref {fig:score_b} are the $D_{\chi^{2}}$ distributions for all events with correct assignment (full red curve) and incorrect assignments (dashed blue curve).

Using the number of events with correct jet-parton assignments in comparison to the total number of events, we find an efficiency of $\varepsilon(t\bar{t}H)=37\%$ to correctly assign all jets to the $t\bar{t}H$ parton final state.
If only the correct reconstruction of the Higgs decay is of interest, the efficiency of the correctly assigned $b$-jets is greater and amounts to $\varepsilon(H)=52\%$ (Table~\ref{tab:efficiency}).

\begin{table}[tbp]
\begin{center}
\begin{tabular}{lcc}
\toprule
Method & $\varepsilon$($t\bar{t}H$) / \% & $\varepsilon$($H$) / \% \\
\midrule
$\chi^2$ method & 37 & 52 \\
Boosted Decision Trees & 45 & 57 \\
Deep Learning & 52 & 63 \\
\bottomrule
\end{tabular}
\caption{Efficiency $\varepsilon(t\bar{t}H)$ of finding the correct jet-parton assignments in $t\bar{t}H$ events, and efficiency $\varepsilon(H)$ of finding the correct jets originating from the Higgs boson decay for all methods.}
\label{tab:efficiency}
\end{center}
\end{table}

In order to obtain more information on the quality of the selected jet-parton assignment, per event we count the permutations with $D_{\chi^{2}}$ value larger than that of the correct assignment $D_{\chi^{2}}^{truth}$ (see Figure~\ref{fig:score_a}):
\begin{align}
I(D)\equiv N_{p}(D>D^{truth})
\label{eq:integral}
\end{align}

The distribution of $I(D_{\chi^{2}})$ can be used to assess the quality of a discriminator.
For correctly selected assignments, $I(D_{\chi^{2}})$ vanishes (full curve in Figure~\ref{fig:score_c}).
However, in the case of an incorrectly selected assignment, $I(D_{\chi^{2}}) > 0$, it should be as small as possible for good discriminators.
For the $\chi^{2}$ method, the resulting distribution of $I(D_{\chi^{2}})$ is shown for all events by the dashed blue curve in Figure~\ref{fig:score_c}.

\subsection{Boosted decision trees}

As an alternative method we use boosted decision trees (BDT) as implemented in \citep{scikit}.
We apply the AdaBoost algorithm for boosting.
With respect to the default hyperparameters we found improved results when using 
an increased number of $400$ decision trees, but no substantial change when varying
the maximum depth of a tree, or the learning rate of the boosting algorithm.
The discriminator $D_{\rm BDT}$ of the BDT is the weighted average of all decision trees each classifying a given jet-parton assignment as signal or background alternatively.

For the BDT we use the same observables as described in section~\ref{sec:observables}.
We apply the same training principle as described in section~\ref{sec:network} and  illustrated in Figure~\ref{fig:architecture_perm}.
For training the BDT, a signal indicator is given in the case that the order of the observables follows the correct jet-parton assignment.
Correspondingly, a background flag is given for other permutations.

On evaluation of the jet-parton assignment in a reconstructed event,  first a preselection is performed using the above-mentioned $\chi^2$ method.
Only the $50$ permutations with the largest $D_{\chi^{2}}$ discriminator (smallest $\chi^2$) values are considered for evaluation by the BDT method.
The permutation with the largest discriminator $D_{\rm BDT}$ is considered as the selected jet-parton assignment.
The efficiency of finding the correct parton-jet assignment is $\varepsilon(t\bar{t}H))=45\%$ which is better compared to the $\chi^2$ method.
Correspondingly, also the efficiency of providing the correct $b$-jets of the Higgs decay is improved: $\varepsilon(H)=57\%$ (see Table~\ref{tab:efficiency}).

As for the $\chi^2$ method, we assess the performance of the BDT discriminator $D_{\rm BDT}$ by examining the number of permutations yielding a higher value than the correct one, $I(D_{\rm BDT})$ (\ref{eq:integral}).
This distribution is shown in Figure~\ref{fig:score_c} by the dashed green curve.
The efficiency of the BDT method is better than that of the $\chi^2$ method as can be seen at $I(D_{\rm BDT})=0$.
Also, the distribution flattens out faster for $I(D_{\rm BDT})>0$ compared to the $\chi^2$ method, indicating a better overall performance of the BDT method.

\subsection{Neural network}

In our third method, we investigate the performance of the neural network.
Network architecture, training and evaluation procedures are described in section~\ref{sec:network}, and the observables in section~\ref{sec:observables}.

For evaluation of the jet-parton assignment in a reconstructed event, 
we again preselect only the $50$ permutations with the largest $D_{\chi^{2}}$ values of the $\chi^2$ method.
On these permutations, the network delivers a corresponding discriminator $D_{\rm DNN}\,\epsilon\,[0,1]$ where the maximum discriminator value leads to the selected jet-parton assignment.
In Figure~\ref {fig:score_d} the distributions of the discriminator $D_{\rm DNN}$ are shown for all events with correct assignment (full red curve) and incorrect assignments (dashed blue curve).

For the majority of the events, the neural network provides the correct jet-parton assignment.
The corresponding efficiency is $\varepsilon(t\bar{t}H))=52\%$, which is better than the BDT method (see Table~\ref{tab:efficiency}).
In $\varepsilon(H)=63\%$ of the events, the network provides the correct jets of the Higgs boson decay.

We also compare the performance of the network by counting permutations with the discriminator $D_{\rm DNN}$ value exceeding that of the correct jet-parton assignment.
The resulting distribution of $I(D_{\rm DNN})$ (\ref{eq:integral}) is shown in Figure~\ref{fig:score_c} for all events by the dashed red curve.
Again, the high efficiency of finding the correct jet-parton assignments is visible at $I(D_{\rm DNN})=0$.
The distribution of the non-zero counts $I(D_{\rm DNN})$ is consistently below the distributions of the two other methods.

\section{\boldmath Impact on $t\bar{t}H$ analyses}

Finding the correct jet-parton assignment in $t\bar{t}H$ events can have a major impact on the sensitivity of a corresponding cross-section analysis.
As a quantitative measure of the improvement in analysis sensitivity is beyond the scope of this work, we will show distributions of key observables typically used in such analyses.

\begin{figure}[t!bp]
\captionsetup[subfigure]{aboveskip=-1pt,belowskip=-1pt}
\begin{centering}
\begin{subfigure}[b]{0.495\textwidth}
\includegraphics[width=\textwidth]{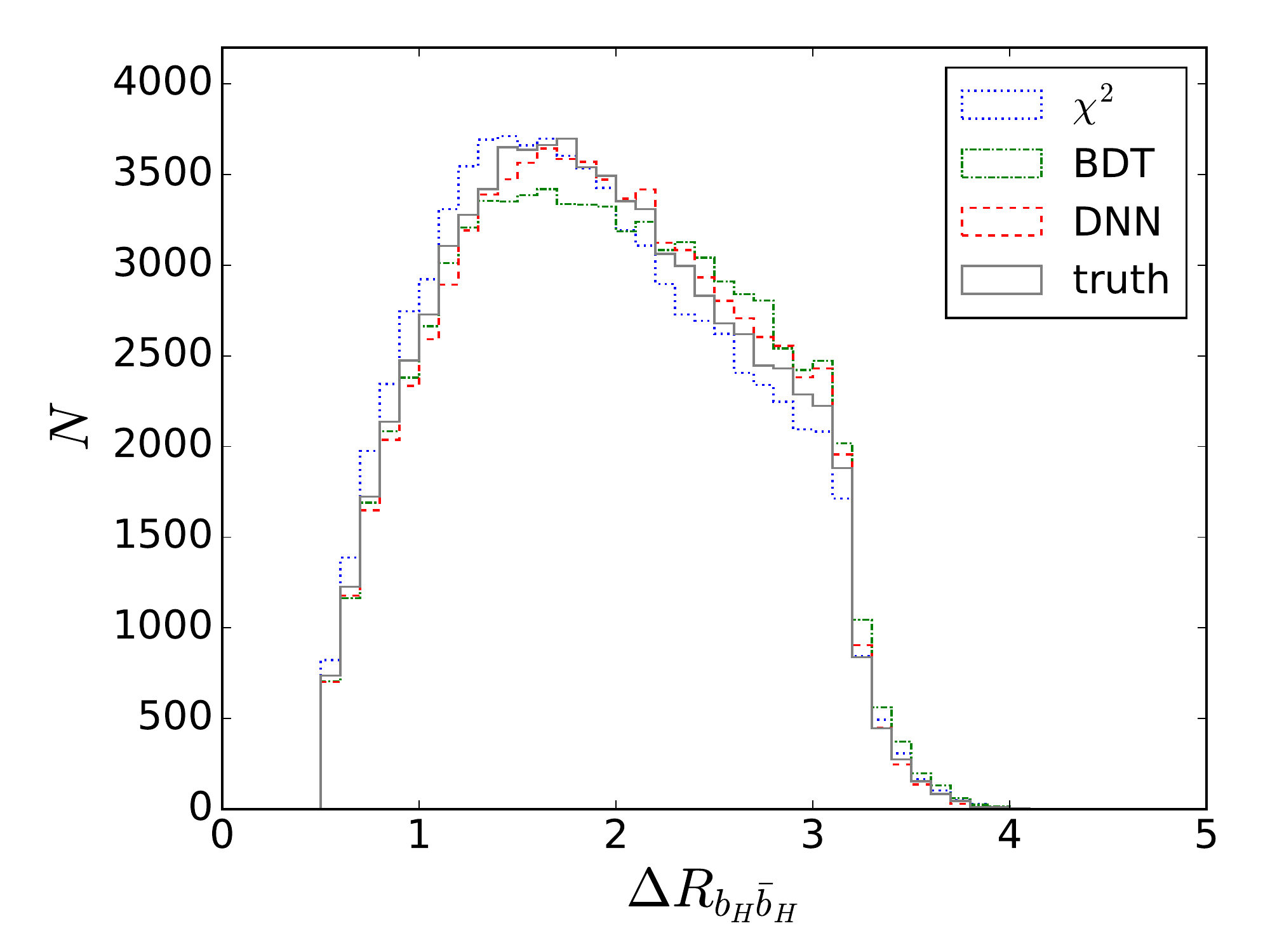}
\subcaption{}
\label{fig:keyplots_a}
\end{subfigure}
\hfill
\begin{subfigure}[b]{0.495\textwidth}
\includegraphics[width=\textwidth]{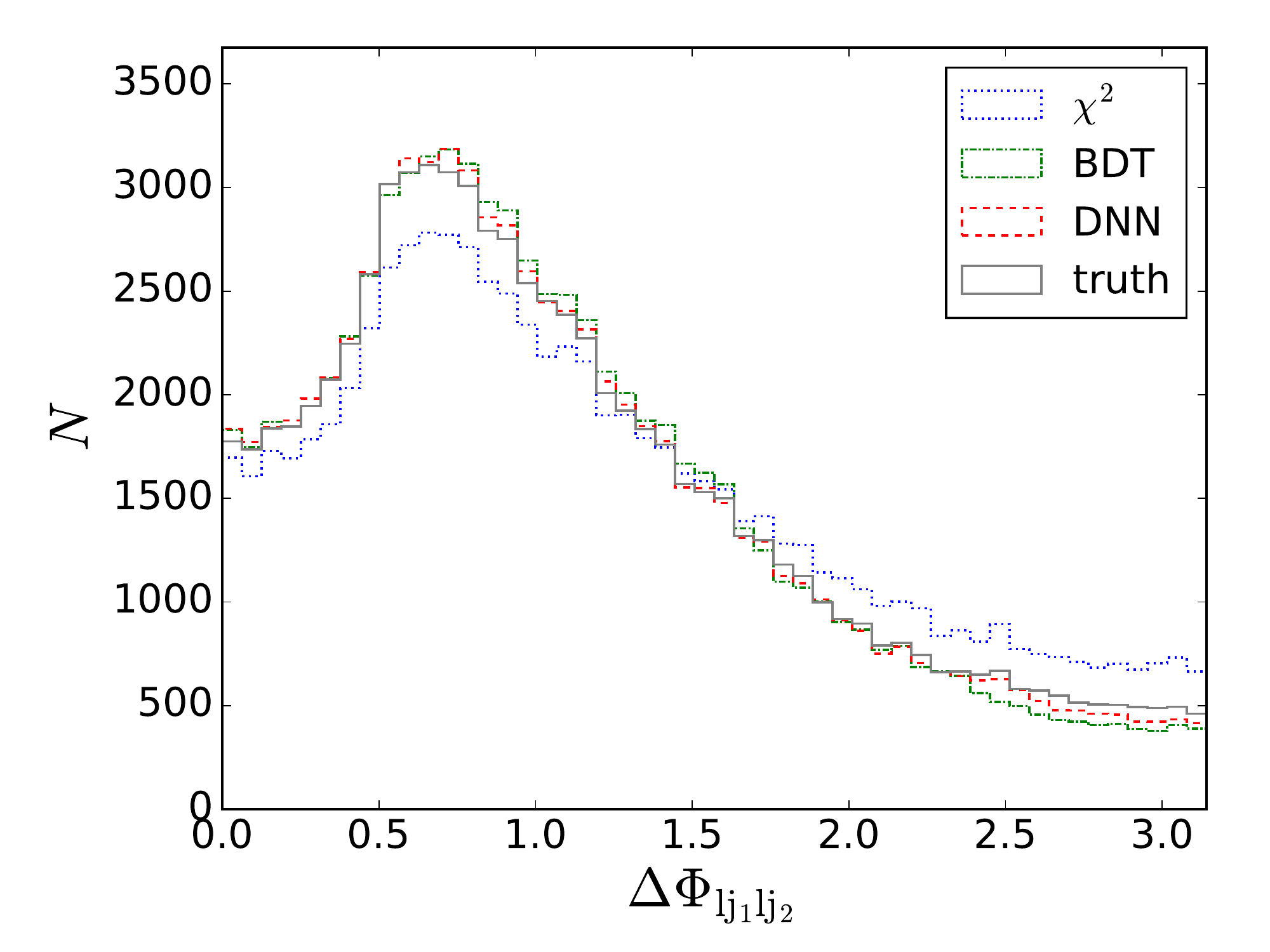}
\subcaption{}
\label{fig:keyplots_b}
\end{subfigure}
\begin{subfigure}[b]{0.495\textwidth}
\includegraphics[width=\textwidth]{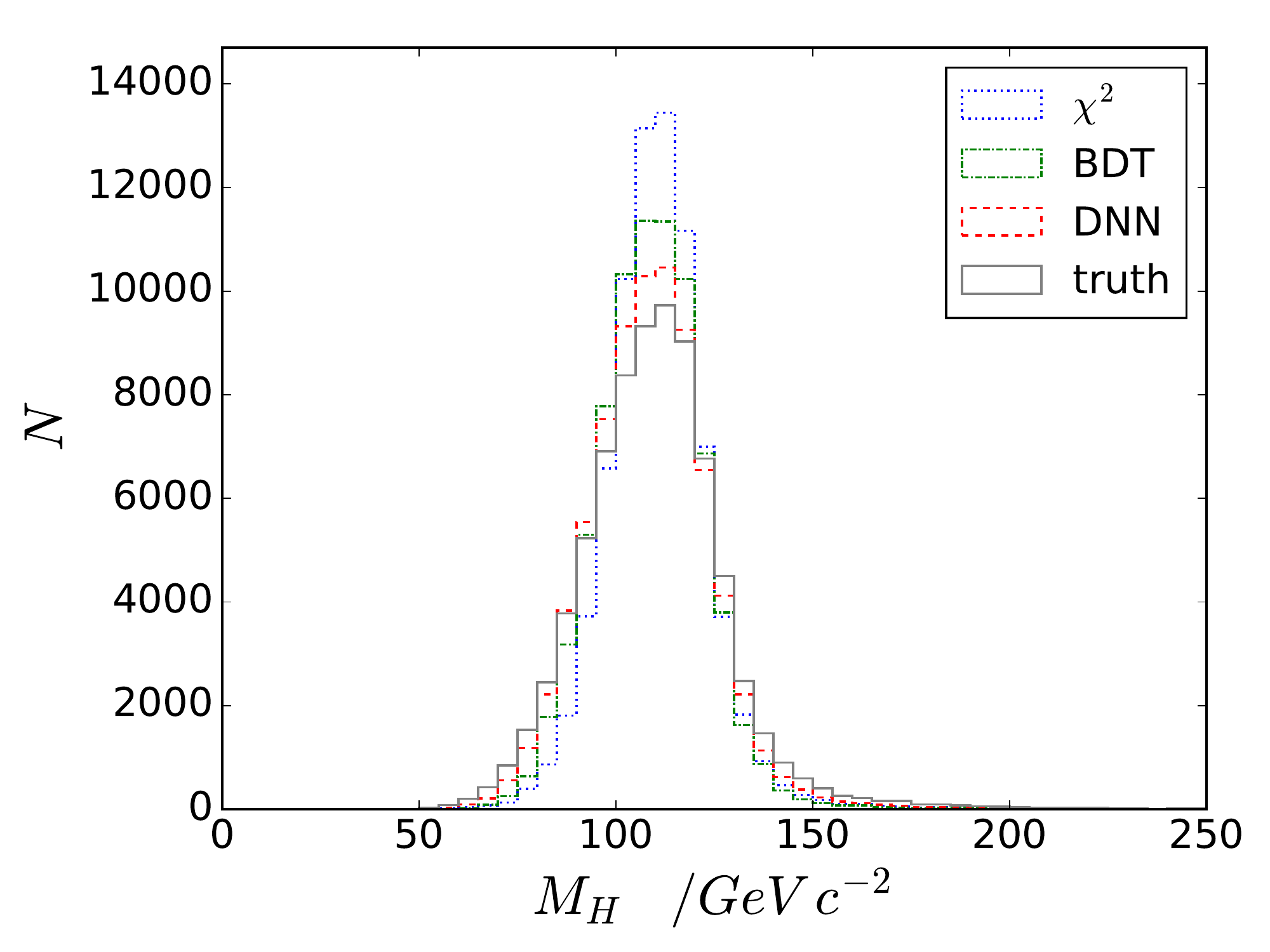}
\subcaption{}
\label{fig:keyplots_c}
\end{subfigure}
\hfill
\begin{subfigure}[b]{0.495\textwidth}
\includegraphics[width=\textwidth]{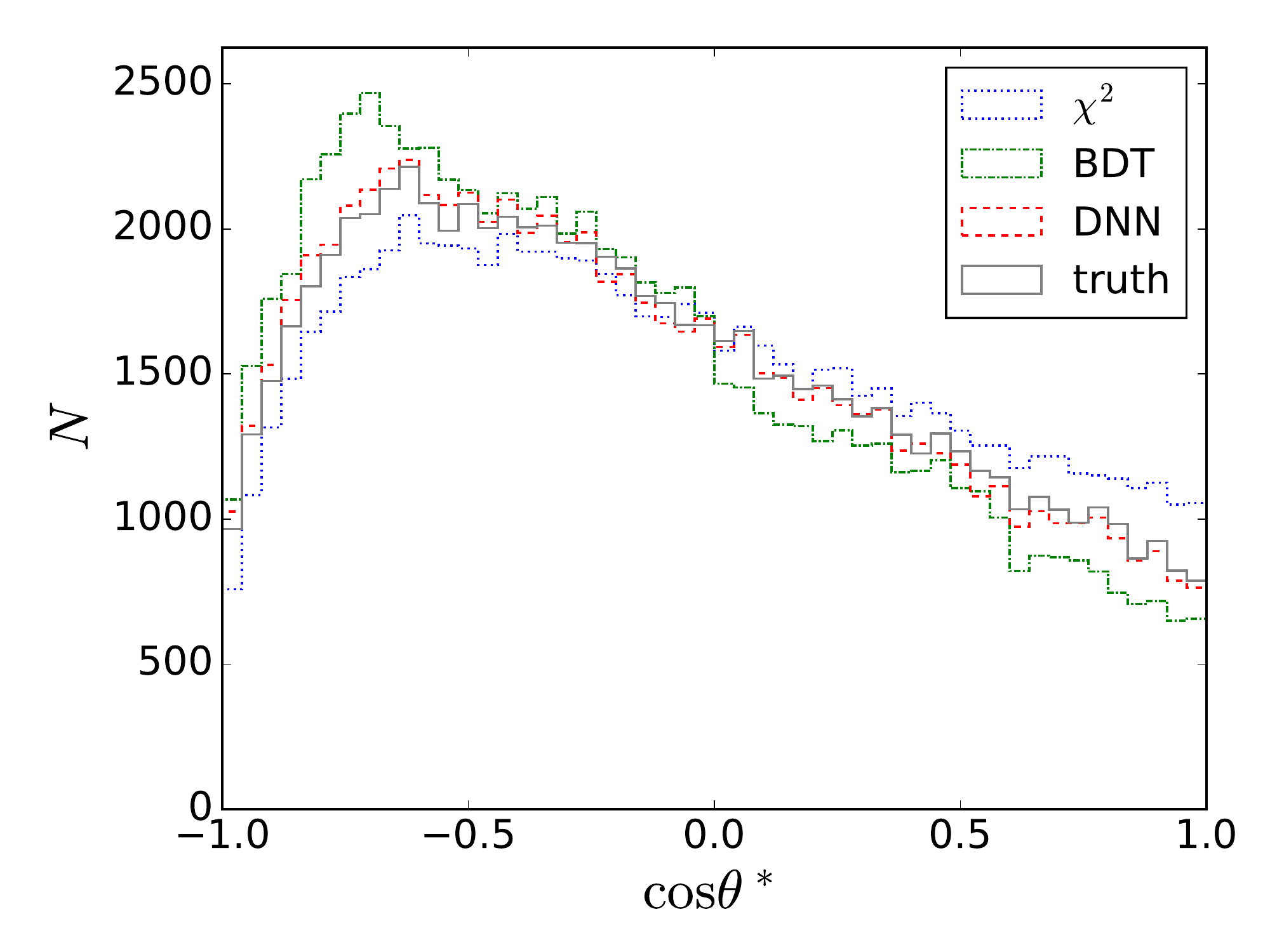}
\subcaption{}
\label{fig:keyplots_d}
\end{subfigure}
\caption{Distributions of reconstructed observables which depend on the correct jet-parton assignments, a) directional distance $\Delta R$ of the $b$-jet pair of the Higgs boson decay, b) azimuthal angular distance of the two jets of the $W$ decay, c) reconstructed Higgs boson mass, and  d) decay angular distribution $\cos{\theta^*}$ of the leptonically decaying $W$ boson.
The full curve shows the correct assignment and the dashed curves depict the neural network (red), the boosted decision tree (green), and the $\chi^{2}$ method (blue).}
\label{fig:keyplots}
\end{centering}
\end{figure}

In Figure~\ref{fig:keyplots_a} we show the distribution of the distance $\Delta R$ of the jets originating from the Higgs boson decay for the correct jet-parton assignment (black curve) in comparison to the three reconstruction methods investigated above.
The result of the neural network is shown by the dashed red curve and consistently depicts the best distribution of correct jet-parton assignments.
The BDT method is illustrated by the dashed green curve which on average yields $\Delta R$ values that are too large.
The $\chi^{2}$ method (dashed blue curve) on average returns $\Delta R$ values that are too small.

Figure~\ref{fig:keyplots_b} shows distributions of the azimuthal angular distance of the two jets of the $W$ boson decay with a similar quality in the description for the BDT and neural network methods but worse for the $\chi^{2}$ method.
The reconstructed mass of the Higgs boson can be seen in Figure~\ref{fig:keyplots_c}.
Both the $\chi^{2}$ and BDT methods have a bias towards the center of the distribution while 
the best possible reconstruction using the correct jet-parton assignment is less pronounced.
The latter distribution is best described by the neural network.

In Figure~\ref{fig:keyplots_d} we also show the angular characteristics of the leptonic $W$ decay.
This is encoded in the distribution of the angle $\theta^*$ between the leptonically decaying $W$ in the rest frame of its top quark, and the charged lepton which is boosted into the $W$ rest frame.
The neural network yields the best description among the three tested methods.

Overall, the neural network provides the best description of the distributions obtained with the correct jet-parton assignment (truth).

\section{Conclusions}

In this work, we investigated the jet-parton assigments in simulated $t\bar{t}H$ events using deep learning techniques.
With a high fraction of correct assignments, high-level variables such as reconstructed masses or decay angular characteristics can be determined more precisely, which is advantageous when separating signal from background processes.

Our study was based on simulated $t\bar{t}H$ events generated with a parameterized detector simulation.
Our investigations were carried out using events which contain all information for reconstructing the $t\bar{t}H$ parton final state.

The fully connected architecture of the neural network contained 8 hidden layers with 500 nodes each and additional connections following the approach of residual networks.
For training, the order of the input variables was according to the correct jet-parton assignment (signal) or a permutation of that order (background), respectively.

During evaluation, a discriminator was calculated for each permutation of the jet-parton assignments, and the assignment with the largest discriminator value was considered the selected assignment which is equivalent to a full event reconstruction.
When compared to two other commonly used methods, the neural network approach returned the best results by far, correctly reconstructing the entire $t\bar{t}H$ event in $52\%$, and only the Higgs boson in $63\%$ of the cases.

\acknowledgments

This work is supported by the Ministry of Innovation, Science and Research of the State of North Rhine-Westphalia, and the Federal Ministry of Education and Research (BMBF).
We wish to thank David Walz for his valuable comments on the manuscript.

\end{document}